\documentclass[conference]{IEEEtran}
\IEEEoverridecommandlockouts
\usepackage{cite}
\usepackage{amsmath,amssymb,amsfonts}

\usepackage{graphicx}
\usepackage{textcomp}

\usepackage{algorithm}
\usepackage{algorithmicx}
\usepackage[noend]{algpseudocode} % no end hides endif endfor

\usepackage{graphicx}
\usepackage{subcaption}

\usepackage{xcolor}
\usepackage{commath} % to use abs

\def\BibTeX{{\rm B\kern-.05em{\sc i\kern-.025em b}\kern-.08em
    T\kern-.1667em\lower.7ex\hbox{E}\kern-.125emX}}

\usepackage{tikz}
\newcommand\copyrighttext{%
  \footnotesize 979-8-3503-1764-0/24/\$31.00~\copyright~2024 IEEE. Personal use permitted. Published in: 2024 IEEE DySPAN. DOI: 10.1109/DySPAN60163.2024.10632739}
\newcommand\copyrightnotice{%
\begin{tikzpicture}[remember picture,overlay]
\node[anchor=south, yshift=15pt] at (current page.south) {\copyrighttext};
\end{tikzpicture}%
}
\begin{document}

\title{3D Spectrum Awareness for Radio Dynamic Zones Using Kriging and Matrix Completion
%\thanks{Identify applicable funding agency here. If none, delete this.}9061157,
\thanks{We use the field measurement data and the preprocessing scripts~\cite{IEEEDataPort_2,IEEEDataPort} published by the NSF Aerial Experimentation and Research Platform on Advanced Wireless (AERPAW)~\cite{9061157} platform. This research is supported in part by the NSF award CNS-1939334.}
}

\author{\IEEEauthorblockN{Mushfiqur Rahman, Sung Joon Maeng, \.{I}smail G\"{u}ven\c{c}, and Chau-Wai Wong}
\IEEEauthorblockA{\textit{Department of Electrical and Computer Engineering, North Carolina State University, Raleigh, NC, USA} \\
%\textit{North Carolina State University}\\
%Raleigh, NC, USA \\
{\tt\small \{mrahman7, smaeng, iguvenc, chauwai.wong\}@ncsu.edu}}
%\and
%\and
%%\textit{name of organization (of Aff.)}\\
%\and
%\and
%\and
}

\maketitle
%\IEEEpubidadjcol
\copyrightnotice

\begin{abstract}
Radio Dynamic Zones (RDZs) are geographically defined areas specifically allocated for testing new wireless technologies. It is essential to safeguard the regular spectrum users outside the zones from the interference caused by the deployed equipment within this zone. Previous works have utilized sparse reference signal received power (RSRP) measurements collected by unmanned aerial vehicles (UAVs) to construct a dense 3D radio map through ordinary Kriging. In this work, we illustrate that matrix completion can outperform ordinary Kriging. We partitioned a 2D area of interest into small square grids where each grid corresponds to a single entry of a matrix. The matrix completion algorithm learns the global structure of the radio environment map by leveraging the low-rank property of propagation maps. Additionally, we illustrate that the simple Kriging and trans-Gaussian Kriging yield better results when the density of known measurements is lower. Earlier works of RSRP prediction involved a training dataset at a single altitude. In this work, we also show that performance can be improved by utilizing a combined dataset from multiple altitudes.
%Furthermore, we demonstrate the advantages in the accumulation of the global structure of radio maps and local spatial correlations by alternating between matrix completion and Kriging interpolation, thereby enhancing performance. 
\end{abstract}


\section{Introduction}
With the rapid development of unmanned aerial vehicles~(UAVs), unmanned ground vehicles (UGVs), smart agriculture devices, and other Internet of things (IoT) technologies, the requirement for continuous and reliable wireless connectivity among the devices is going to be higher in the future. The throughput demand for smartphones is also increasing due to high-resolution videos and the surge of smartphone applications. One way of mitigating the huge demand for wireless communications is to share the spectrum among various devices or services. Innovative wireless technologies such as spectrum sharing can be deployed and tested within a radio dynamic zone (RDZ)~\cite{maeng2022national, zheleva2023radio,tschimben2023testbed}. While developing spectrum-efficient technologies using necessary wireless devices inside the zone, RDZ intends to isolate its radio environment from the rest of the world. In particular, inference management near the boundary of an RDZ is not an easy task and requires deploying spectrum monitoring sensors within the zone.

Creating a dense radio environment map from a monitoring system that involves sparse measurement sensors poses a significant challenge, making it an active area of research. Kriging~\cite{sato2017Kriging,braham2016fixed,yilmaz2013radio} based algorithm has been proposed for interpolating at an unknown location using a linear combination of the known reference signal received power (RSRP) values exploring spatial correlation. Several studies~\cite{graziosi2002general, szyszkowicz2010feasibility, simunek2013uav} have focused on modeling the spatial correlation among various points. Kernel-based radio environment mapping~\cite{teganya2019location, bazerque2013nonparametric, hamid2017non} predicts the RSRP values as a combination of kernel functions. Matrix completion~\cite{migliore2011compressed, sun2021grid, zhang2020spectrum} is also used for radio environment mapping. It divides a region into a spatial grid and puts data into a matrix. Known sparse data points are used to fill the nearby unknown entries leveraging the low-rank property.

\begin{figure}[!t]
\centerline{\includegraphics[width=\linewidth,trim={0.5cm 1cm 1cm 2.2cm},clip]{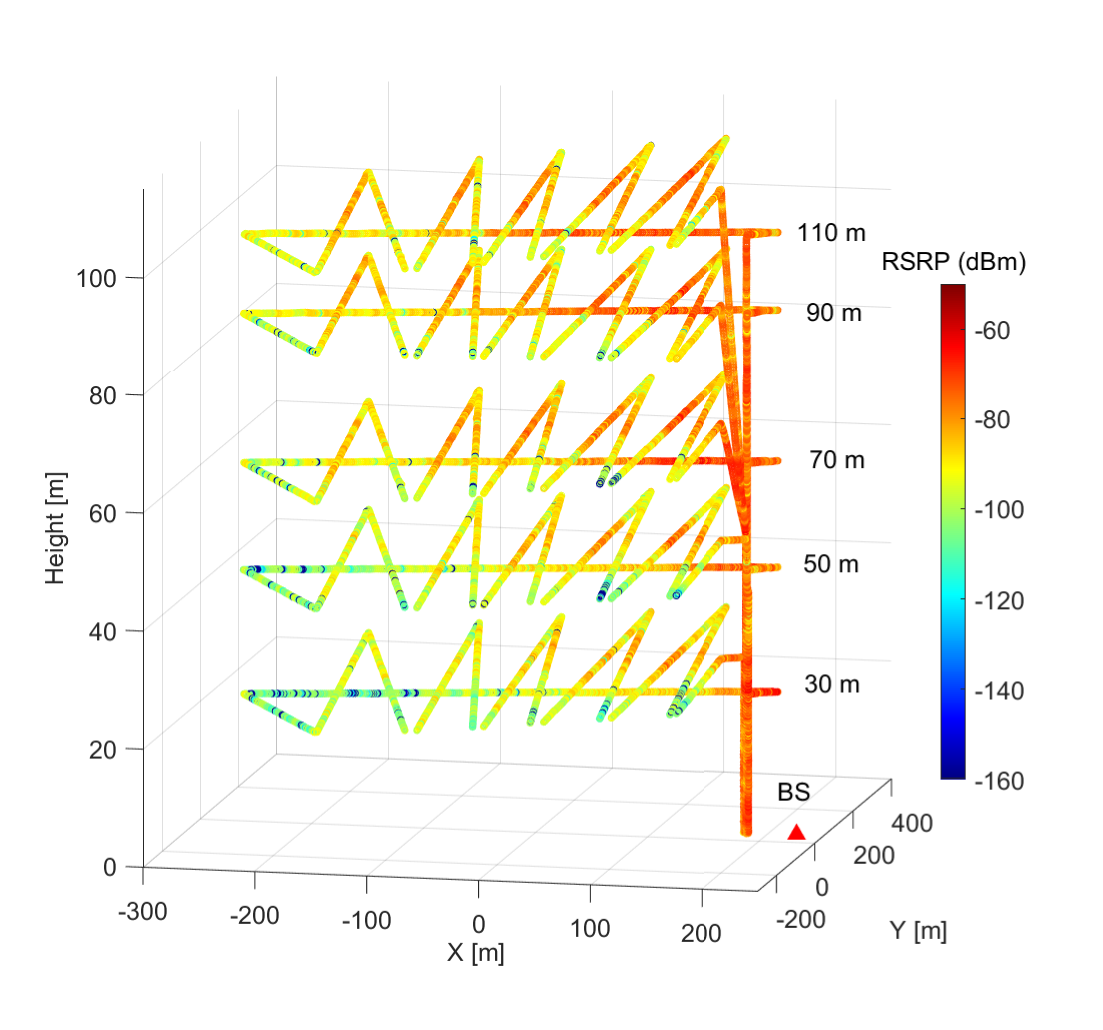}}
\caption{3D sparse RSRP measurement dataset~\cite{IEEEDataPort_2} used in this study. The measurements were collected by a UAV that aimed to follow a predefined zig-zag pattern at five different altitudes.}
\label{fig:uav_trajectory}
\vspace{-2mm}
\end{figure}
Maeng et al.~\cite{maeng2022out} proposed Kriging interpolation for spectrum leakage estimation at the boundary of circular-shaped RDZ using simulated data. In more recent work, Maeng et al.~\cite{maeng2023kriging} utilized Kriging interpolation with a two-ray pathloss model, shadow fading, and measured antenna radiation pattern. A real dataset was used to test the performance of Kriging interpolation for generating a 3D radio environment map. The dataset~\cite{IEEEDataPort_2} collected by the AERPAW team contains RSRP measurements of 5 different altitudes, ranging between $30$~m and $110$~m. The signal strengths were measured by a UAV that followed a zig-zag trajectory while measuring RSRPs by the mounted sensors. Fig.~\ref{fig:uav_trajectory} shows all the sparse measurements available in the dataset. In this study, we use the same dataset to conduct a comprehensive analysis of performance for matrix completion and various types of Kriging interpolation schemes. Our contributions are as follows:

\begin{itemize}
    \item We demonstrate that matrix completion can outperform Kriging interpolation for dense spectrum awareness.% and propose an algorithm that allows switching between Kriging and matrix completion to effectively combine the advantages of the two heterogeneous algorithms.
    \item We show that simple Kriging can significantly outperform ordinary Kriging at a low-data regime, and trans-Gaussian Kriging can further improve the performance.
    \item We show via experiments that training data from multiple heights can improve the interpolation performance.
\end{itemize}
The rest of the paper is organized as follows. Section~\ref{sec:sys_model} establishes the base of our analysis. Section~\ref{sec:kriging} and Section~\ref{sec:mat_completion} present Kriging interpolation and matrix completion using single height measurements. Section~\ref{sec:num_results} presents the numerical results and the last section concludes our paper.

\section{System Model Preliminaries}
\label{sec:sys_model}
\subsection{Radio Propagation Model}
Let us define the location of a base station $l^{\mathrm{bs}} = \left(\psi^{\mathrm{bs}}, \zeta^{\mathrm{bs}}, h^{\mathrm{bs}}\right)$ and the location of the UAV $l^{\mathrm{uav}} = \left(\psi^{\mathrm{uav}}, \zeta^{\mathrm{uav}}, h^{\mathrm{uav}}\right)$, where $\psi$, $\zeta$ and $h$ denote the latitude, longitude, and height correspondingly. The horizontal and vertical distances between the base station and the UAV are:
\begin{subequations}
\begin{align}
d_{\mathrm{h}}\left(l^{\mathrm{bs}}, l^{\text {uav }}\right) & = A \cdot \arccos \Big(\sin \psi^{\text {uav }} \sin \psi^{\mathrm{bs}}\Big. \nonumber\\
&\Big.+\cos \psi^{\text {uav }} \cos \psi^{\mathrm{bs}} \cos (\zeta^{\mathrm{bs}}-\zeta^{\text {uav }})\Big) \label{eqn:dh},\\
d_{\mathrm{v}}\left(l^{\mathrm{bs}}, l^{\text {uav }}\right) & =\big|h^{\mathrm{bs}}-h^{\text {uav }}\big|,
\end{align}
\end{subequations}
where $A$ is the radius of the earth. The 3D distance, $d_{3\text{D}}(l^{\mathrm{bs}}, l^{\mathrm{uav}})$ can be computed as $\sqrt{d_\text{h}^2+d_\text{v}^2}$.

\begin{figure}% 
    \centering
		\vspace{-0mm}
    \includegraphics[width=\linewidth,trim={4.3cm 8.2cm 11.8cm 7.5cm},clip]{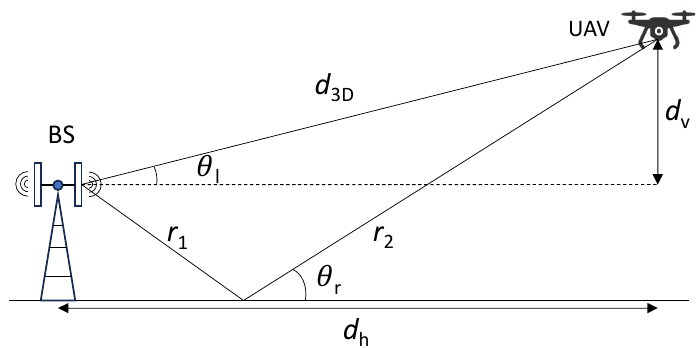}
    \hspace{4mm}
    \vspace{-3mm}
    \caption{Signal propagation model used in this study involves a line of sight (LOS) and a ground reflected path from a base station to a UAV. Pathloss depends on $d_\text{3D},r_\text{1},r_\text{2}, \theta_\text{l},$ and $\theta_\text{r}$.}
	\vspace{-2mm}
    \label{fig:diagram_two-ray-model}
    \vspace{-1mm}
\end{figure}

The two-ray ground reflection model is illustrated in Fig.~\ref{fig:diagram_two-ray-model}.
The received signal has two components, one from a line of sight (LoS) path and the other from a strong ground reflection path. The pathloss for the combined signal is given as follows:
\begin{equation}
\begin{aligned}
\mathrm{PL}_{\text {twm }}&\left(l^{\mathrm{bs}}, l^{\text {uav }}\right)=\left(\frac{\lambda}{4 \pi}\right)^2  \, \biggr \lvert \underbrace{\frac{\sqrt{\mathrm{G}_{\mathrm{bs}}\left(\phi_\text{l}, \theta_\text{l}\right) \mathrm{G}_{\mathrm{uav}}\left(\phi_\text{l}, \theta_\text{l}\right)}}{d_{3 \mathrm{D}}}}_{\text {LoS signal }}\\
& +\underbrace{\frac{\Gamma\left(\theta_\text{r}\right) \sqrt{\mathrm{G}_{\mathrm{bs}}\left(\phi_\text{r}, \theta_\text{r}\right) \mathrm{G}_{\mathrm{uav}}\left(\phi_\text{r}, \theta_\text{r}\right)} e^{-j \Delta \tau}}{r_1+r_2}}_{\text {ground reflected signal }}\biggr\rvert^2,
\end{aligned}
\label{eqn:two-ray-model}
\end{equation}
where $G_{\text{bs}}(\theta,\phi)$, $G_{\text{uav}}(\theta,\phi)$ are the antenna gain of the base station and the UAV, $\theta$ is the elevation angle, $\phi$ is the azimuth angle, $\theta_\text{l}$ and $\theta_\text{r}$ are the two angles as shown in Fig.~\ref{fig:diagram_two-ray-model}, $\lambda$ is the wavelength, $\Delta \tau$ is phase delay between two paths, and $\Gamma\left(\theta_\text{r}\right)$ is the ground reflection coefficient. The overall received power signal is given by:
\begin{equation}
r=\mathrm{P}_{\mathrm{Tx}}-\mathrm{PL}_{\mathrm{twm}}^{(\mathrm{dB})}+w,
\end{equation}
where $\mathrm{P}_{\mathrm{Tx}}$ is the transmit power, $\mathrm{PL}_{\mathrm{twm}}^{(\mathrm{dB})}$ is the pathloss from~\eqref{eqn:two-ray-model}, and $w$ is the shadow fading, all in the dB scale.

\subsection{Antenna Radiation Pattern}
For typical air-to-ground communication scenarios, antenna gain is sensitive to elevation angle but remains unchanged by the azimuth angle~\cite{maeng2023kriging}. The dataset~\cite{IEEEDataPort_2} considered for this work used a receiver dipole antenna (SA-1400-59000) for RSRP measurement that was attached to the UAV. The receive antenna has a donut-shaped radiation pattern as specified in the vendor's sheet~\cite{sa_1400}. This follows an omnidirectional pattern along the azimuth angle and an oval-shaped pattern in the elevation angle. The radiation pattern is not sensitive to the frequency as well. The transmit dipole antenna,  RM-WB1-DN, on the other hand, is not guaranteed to follow omnidirectionality~\cite{mobilemarkrmwb1} along the azimuth angle. The specification sheet also does not provide the radiation pattern for the frequency used in the experiment. For transmit antenna gain, we used the published 3D antenna radiation pattern~\cite{IEEEDataPort} of RM-WB1-DN that was measured in an anechoic chamber.
%To overcome the limitations of the inaccurate radiation patterns, we used the 3D antenna radiation pattern measured in an anechoic chamber facility at Wireless Research Center (WRC), Wake Forest, NC.

\subsection{Correlation Modeling In 3D space}
\begin{figure}% 
    \centering
		\vspace{-0mm}
    \includegraphics[width=\linewidth,trim={0.5cm 1.0cm 0.5cm 2.6cm},clip]{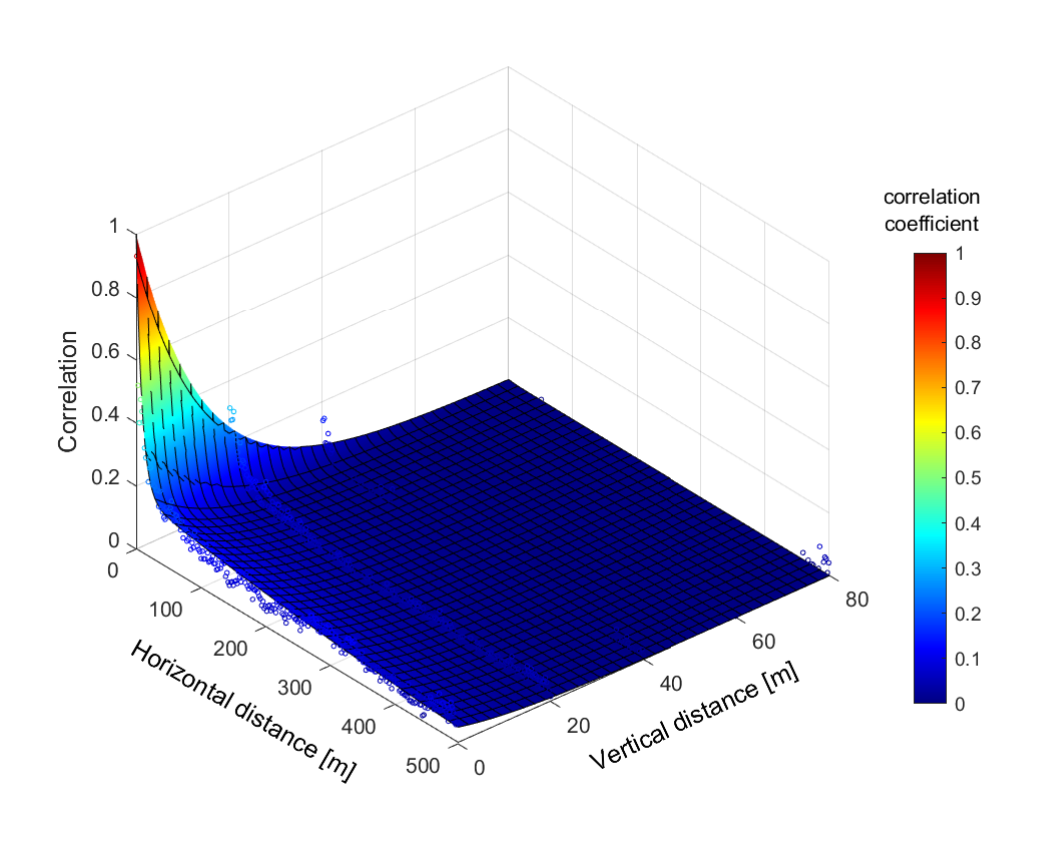}
    \hspace{4mm}
    \vspace{-3mm}
    \caption{Correlation modeling in a 3D space considering the horizontal and vertical distances. The correlation model serves as an important characteristic for interpolation.}
	\vspace{-2mm}
    \label{fig:correlation-model}
    \vspace{-1mm}
\end{figure}

The received signal power between two different locations typically follows a spatial correlation based on the distance~\cite{graziosi2002general}. The spatial correlation of the received signal mainly originates from the spatial correlation of the shadow-fading component~\cite{szyszkowicz2010feasibility}. The spatial correlation between the two UAV locations $l_i^{\mathrm{uav}}$ and $l_j^{\mathrm{uav}}$ in a 3D space can be modeled as a function of horizontal and vertical distances~\cite{maeng2023kriging}:
\begin{equation}
R\left(l_i^{\text {uav }}, l_j^{\text {uav }}\right)=\frac{\mathbb{E}\left[w(l_i^{\text {uav }}) w(l_j^{\text {uav }})\right]}{\sigma_w^2}=R(d_{\mathrm{v}}, d_{\mathrm{h}}),
\end{equation}
where $\sigma_w^2$ is the variance of shadow fading. In this study, we model the 3D correlation combining the biexponential decay function of the horizontal distance and the exponential decay function of the vertical distance as follows:
\begin{equation}
\label{eqn:corr_model}
R\left(d_{\mathrm{v}}, d_{\mathrm{h}}\right)=e^{-qd_{\mathrm{v}}}[ae^{-p_1d_{\mathrm{h}}}+(1-a)e^{-p_2d_{\mathrm{h}}}],
\end{equation}
where $a$, $p_1$, $p_2$, and $q$ are obtained through fitting with the field measurement data. The fitted model is shown in Fig.~\ref{fig:correlation-model}.
Note that we chose this model as it was found~\cite{maeng2023kriging} that the correlation model fits better with an exponential function for the vertical distance and with a biexponential function for the horizontal distance.

\section{Kriging Interpolation}
\label{sec:kriging}
In this section, we present three different variants of Kriging interpolation: ordinary Kriging~\cite{wackernagel2003ordinary}, simple Kriging~\cite{olea1999simple}, and trans-Gaussian Kriging~\cite{cressie2015statistics}.

\subsection{Ordinary Kriging}
\label{subsection:ordinary_Kriging}
%Given a set of known measurements $\{Z(s_1),...,Z(s_n)\}$ at $n$ spatial locations $s_1,...,s_n$, ordinary Kriging uses a linear combination of known measurements to predict the measurement $Z$ at any arbitrary location $s_0$. The goal is to minimize the square of the estimation error while restricting the sum of weights for linear combination to $1$.
Let us consider known signal measurements, i.e., RSRP data denoted as ${Z(s_1), \dots, Z(s_n)}$, collected at distinct spatial locations $s_1, \dots, s_n$, where $s$ is an alternate notation of $l^{\mathrm{uav}} = \left(\psi^{\mathrm{uav}}, \zeta^{\mathrm{uav}}, h^{\mathrm{uav}}\right)$. In wireless spectrum awareness, these data can be collected either by placing $n$ receivers at different spatial locations or maneuvering a UAV equipped with a receiver and collecting the data during the mission duration. Ordinary Kriging uses a linear combination of these known measurements to predict the signal $\hat{Z}(s_0)$ at given location $s_0$. The goal is to minimize the square of the estimation error while ensuring that the sum of the weights of the linear combination adds up to 1. The objective of ordinary Kriging can be formally written as:

\begin{equation}
\label{eqn:ordinary_kriging_inference}
\begin{aligned}
\min _{\lambda_{1}, \ldots, \lambda_{n}} & \mathbb{E}\left[\left(\hat{Z}\left(s_0\right)-Z\left(s_0\right)\right)^2\right], \\
\text { s.t. } & \hat{Z}\left(s_0\right)=\sum_{i=1}^n  \lambda_{i} Z(s_i), \\
& \sum_{i=1}^n \lambda_{i}=1.
\end{aligned}
\end{equation}
This is a constrained minimization problem and can be solved with the Lagrange optimization technique. The solution may be obtained by solving the following set of linear equations:

\begin{equation}
\begin{aligned}
& {\left[\begin{array}{cccc}
\mathbf{\gamma}(s_1,s_1) & \cdots & \mathbf{\gamma}\left(s_1,s_n\right) & 1 \\
%\mathbf{\gamma}(s_2,s_1) & \cdots & %\mathbf{\gamma}\left(s_2,s_n\right) & 1 \\
\vdots & \ddots & \vdots & \vdots \\
\mathbf{\gamma}(s_n,s_1) & \cdots & \mathbf{\gamma}\left(s_n,s_n\right) & 1 \\
1 & \cdots & 1 & 0
\end{array}\right]\left[\begin{array}{c}
\lambda_{1} \\
%\lambda_{2} \\
\vdots \\
\lambda_{n} \\
\mu
\end{array}\right]} \\
& =\left[\begin{array}{c}
\mathbf{\gamma}\left(s_0,s_1\right) \\
%\mathbf{\gamma}\left(s_0,s_2\right) \\
\vdots \\
\mathbf{\gamma}\left(s_0,s_n\right) \\
1
\end{array}\right] ,
\end{aligned}
\label{eq:ordinary_kirging_soln}
\end{equation}
where $\mathbf{\gamma}\left(s_i,s_j\right)$ denotes the semivariogram~\cite{olea1994fundamentals} between two spatial locations $s_i$ and $s_j$ with respect to $Z$ given by:
\begin{equation}
\mathbf{\gamma}\left(s_i,s_j\right)  = \mathrm{Var}\big(Z(s_i)-Z(s_j)\big),
\end{equation}
and $\mu$ is the Lagrange multiplier. The analytical expression of the mean-squared error (MSE) for prediction at $s_0$ is given by
\begin{equation}
\label{eqn:ordianry_Kriging_variance}
\sigma_{\text{ok}}^2\left(s_0\right) = \sum_{i=1}^n \lambda_i \gamma\left(s_0,s_i\right)+\mu.
\end{equation}
%Denoting the known measurements as $\boldsymbol{M}_{\text{Z}}$, corresponding latitudes $\boldsymbol{M}_{\psi}$ and corresponding longitudes $\boldsymbol{M}_{\zeta}$, Algorithm~\ref{alg:mc_kriging} outlines the scheme of ordinary Kriging.
%\begin{algorithm}[!t]
    %\caption{Interpolation using ordinary Kriging}
    %\label{alg:mc_kriging}
        %\begin{algorithmic}[1]
            %\Procedure{$\textnormal{OrdinaryKriging}$}%{}%\Comment{The g.c.d. of a and b}
            %\State $\textbf{Input:}$
            %\State $\boldsymbol{M}_{\psi} \in \mathbb{R}^{N_k}, \boldsymbol{M}_{\zeta} \in \mathbb{R}^{N_k}$%\Comment{$s_1,\dots,s_n$}
            %\State $\boldsymbol{M}_{\text{Z}} \in \mathbb{R}^{N_k}$%\Comment{$Z(s_1),\dots,Z(s_n)$}
            %\State $\psi_{\text{0}} \in \mathbb{R}, \zeta_{\text{0}} \in \mathbb{R}$%\Comment{$s_0$}
            %\State \textbf{Compute:} $\hat{Z}_{\text{ok}}(\psi_\text{0}, \zeta_\text{0})$ , $ \sigma^2_{ok}(\psi_\text{0}, \zeta_\text{0})$ from \eqref{eq:ordinary_kirging_soln} and \eqref{eqn:ordianry_Kriging_variance}
            %\State $\textbf{Output:}\,\hat{Z}_{\text{ok}}(\psi_\text{0}, \zeta_\text{0})\,,\sigma^2_{\text{ok}}(\psi_\text{0}, \zeta_\text{0})$
            %\EndProcedure
        %\end{algorithmic}
%\end{algorithm}
\subsection{Simple Kriging}
Simple Kriging, also known as the best linear predictor, estimates a measurement at a spatial location $s_0$ from a linear combination of the known measurements ${Z(s_1), ..., Z(s_n)}$ within the boundary region. The assumption is that the mean $m_Z$ of measurement variable $Z$ and the covariance between the samples of $Z$ measured at spatial locations $s_i$ and $s_j$, are known. The covariance decays exponentially with respect to spatial distance, which can be reformulated to the semivariogram as follows:

\begin{equation}
C(s_i, s_j) = \sigma^2_Z - \gamma(s_i, s_j),
\end{equation}
where $\sigma^2_Z$ is the variance of the measurement variable $Z$. The objective is again to minimize the MSE at any arbitrary location $s_0$ as follows:

\begin{eqnarray}
\label{eqn:simple_kriging_inference}
\min _{\lambda_{1}, \ldots, \lambda_{n}} & \mathbb{E}\left[\left(\hat{Z}\left(s_0\right)-Z\left(s_0\right)\right)^2\right], \nonumber \\
\text { s.t. } & \hat{Z}\left(s_0\right)=m_Z + \sum_{i=1}^n  \lambda_{i} \left(Z\left(s_i\right) - m_Z\right).
\end{eqnarray}
The optimum weights for simple Kriging can be computed by solving the following linear equation:
\begin{equation}
\label{eqn:simple_Kriging_soln}
\begin{aligned}
& {\left[\begin{array}{cccc}
C(s_1,s_1) & \cdots & C\left(s_1,s_n\right) \\
%C(s_2,s_1) & \cdots & C\left(s_2,s_n\right) \\
\vdots & \ddots & \vdots \\
C(s_n,s_1) & \cdots & C\left(s_n,s_n\right)
\end{array}\right]\left[\begin{array}{c}
\lambda_{1} \\
%\lambda_{2} \\
\vdots \\
\lambda_{n} \\
\end{array}\right]} \\
& =\left[\begin{array}{c}
C\left(s_0,s_1\right) \\
%C\left(s_0,s_2\right) \\
\vdots \\
C\left(s_0,s_n\right)
\end{array}\right] .
\end{aligned}
\end{equation}
The estimated prediction MSE for simple Kriging is
\begin{equation}
\label{eqn:simple_Kriging_variance}
\sigma_\text{sk}^2\left(s_0\right) = \sigma_Z^2 - \sum_{i=1}^n \lambda_i C\left(s_0,s_i\right).
\end{equation}

\subsection{Trans-Gaussian Kriging}
%The performance of Kriging interpolation depends on the reliability of the semivariogram and the fulfillment of stationary property of data in the spatial axes. If the measurement $Z$ is not intrinsically stationary in one domain, it is intuitive to transform the data in another domain $Y$ where the data is stationary. 
We noticed that our computed shadow-fading data based on field measurements has a skewed distribution. We know that Gaussian distribution ensures the optimality of a linear optimal predictor over all possible predictors including nonlinear ones in terms of minimizing MSE~\cite[Chapter 3]{cressie2015statistics}.
%We found our computed shadow-fading data based on field measurements has a skewed Gaussian distribution, where we know that the shadow-fading data is stationary and follows a Gaussian distribution. 
Hence, we transform our shadow-fading data $Z$ into $Y$ that has a standard normal distribution $\mathcal{N}(0,1)$. The transformation function $f(\cdot)$ can be obtained as follows~\cite{casella2021statistical}:
\begin{equation}
Y(s) = \operatorname{CDF}_{\mathcal{N}(0,1)}^{-1}\Big(\operatorname{CDF}_{Z}\big(Z(s)\big)\Big) \equiv f\big(Z(s)\big),
\end{equation}
where $\operatorname{CDF}(\cdot)$ is the cumulative distribution function (CDF) of a random variable.
Let us denote the mean of $Y$ as $m_Y$ and the variance of $Y$ as $\sigma^2_Y$, the inverse function of $f(\cdot)$ as $\phi(\cdot)$, the ordinary Kriging prediction of $Y$ as $\hat{Y}_\text{ok}(\cdot)$, the simple Kriging prediction of $Y$ as $\hat{Y}_\text{sk}(\cdot)$, their corresponding MSE calculated by~\eqref{eqn:ordianry_Kriging_variance},~\eqref{eqn:simple_Kriging_variance} as $\sigma_{Y,\text{ok}}^2\left(\cdot\right)$ and $\sigma_{Y,\text{sk}}^2\left(\cdot\right)$, and the lagrange multiplier in \eqref{eq:ordinary_kirging_soln} as $\mu_Y$.

After predicting measurement $Y$ at any arbitrary location $s_0$, we need to transform it back into the $Z$ domain using $\phi(\cdot)$. Moreover, it turns out that estimation in the $Z$ domain has a prediction bias if the second-order derivative of $\phi(\cdot)$ computed at $m_Y$ is non-zero. To compensate for the bias, the approximately unbiased solution for ordinary Kriging is given as follows~\cite{cressie2015statistics}:
\begin{equation}
\hat{Z}_\text{ok}\!\left(s_0\right)=\phi\!\left(\hat{Y}_\text{ok}(s_0)\right)+\phi^{\prime \prime}\!\left(m_Y\right)\left[\sigma_{Y, \text{ok}}^2\left(s_0\right) / 2\!-\!\mu_Y\right].
\end{equation}
The above equation originates from the second-order delta method.
Similarly, for simple Kriging, the approximately biased solution is given by~\cite{cressie2015statistics}:
\begin{equation}
\begin{aligned}
\hat{Z}_\text{sk}\!\left(s_0\right)&=\phi\left(\hat{Y}_\text{sk}(s_0)\right) \\
&+\frac{\phi^{\prime \prime}\!\left(m_Y\right)} {2}\left[\sigma_{Y,\text{sk}}^2\left(\mathbf{s}_0\right)-\sum_{i=1}^n \lambda_i C\left(s_0,s_i\right)\right].
\end{aligned}
\end{equation}

\subsection{3D Interpolation}
\label{sec:3d_interp}
While RSRP interpolation aims to create a radio environment map of a 3D volume from sparse and incomplete measurements, it is not possible to obtain measurements at each altitude and over a dense grid. For example, suppose we use a UAV to collect measurements while the UAV is covering an area at a constant height. In this case, the collected data is associated with a certain height from the ground. For 3D awareness of radio signal coverage, it might be necessary not only to interpolate in the horizontal plane of the UAV trajectory in which data were collected but also to predict the signal strength at other heights that the UAV did not cover. In this study, we additionally utilize the measured RSRP values datasets at other altitudes and compare the performance with the regular same-height prediction. Moreover, while using the same-height dataset, we increase the number of known observations with the measurements at different altitudes with $20$~m or $40$~m vertical distance, which will be shown to improve the performance.

\section{Matrix Completion for Spectrum Awareness}
\label{sec:mat_completion}
Given a collection of sparse RSRP measurements, the matrix completion based interpolation involves three steps. First, the sparse measurements are converted into a 2D matrix containing both known and unknown entries. Second, the unknown matrix entries are populated by adjusting the entire matrix. Third, for any arbitrary coordinate, RSRP is predicted from the completed matrix.
Denoting the known measurements as $\boldsymbol{M}_{\text{Z}}$, corresponding latitudes $\boldsymbol{M}_{\psi}$, corresponding longitudes $\boldsymbol{M}_{\zeta}$, prediction latitude ${\psi_0}$, and prediction longitude ${\zeta_0}$,
the detailed interpolation algorithm is given in Algorithm~\ref{alg:mc_algo}.

\begin{algorithm}[!t]
    \caption{Interpolation of sparse RSRP data using matrix completion}
    \label{alg:mc_algo}
        \begin{algorithmic}[1]
            \Procedure{$\textnormal{MatrixCompletion}$}{}%\Comment{The 
            \State $\,\textbf{Input:} \boldsymbol{M}_{\text{Z}} \in \mathbb{R}^{N}, \boldsymbol{M}_{\psi} \in \mathbb{R}^{N}, \boldsymbol{M}_{\zeta} \in \mathbb{R}^{N}, \psi_{\text{0}} \in \mathbb{R}, \zeta_{\text{0}} \in \mathbb{R}$
            \State $\textbf{Initialization:}$
            \State  $\text{d}_{\text{grid}} \gets $ Constant, equivalent to 5\,m
            \State  $\boldsymbol{G}_{\psi} \gets \text{min}(\boldsymbol{M}_{\psi}):\text{d}_{\text{grid}}:\text{max}(\boldsymbol{M}_{\psi})$
            \State  $\boldsymbol{G}_{\zeta} \gets \text{min}(\boldsymbol{M}_{\zeta}):\text{d}_{\text{grid}}:\text{max}(\boldsymbol{M}_{\zeta})$
            \State $N_\text{r} = \text{count}(\boldsymbol{G}_{\psi})$, $N_\text{c} = \text{count}(\boldsymbol{G}_{\zeta})$
            \State  $\boldsymbol{S}_{i} \gets \{1,2,\dots,N_{\text{r}}\}$, $\boldsymbol{S}_{j} \gets \{1,2,\dots,N_{\text{c}}\}$ 
            \State $\boldsymbol{Z_{\text{}}} \in \mathbb{R}^{N_{\text{r}} \times N_{\text{c}}}$ \Comment{RSRP}
            \State $\boldsymbol{\Sigma_{\text{}}} \in \mathbb{R}^{N_{\text{r}} \times N_{\text{c}}}$ \Comment{MSE}
            \State $\textbf{I}_\Omega \gets (0)^{N_{\text{r}} \times N_{\text{c}}}$ \Comment{Known indicator}
            \State \textbf{Local interpolation:}
            \State $\boldsymbol{S_{\text{m}}}=\{1,\dots,20\}$
            \For{$i \in \boldsymbol{S}_{i}$}
            \For{$j \in \boldsymbol{S}_{j}$}
            \State $\boldsymbol{d_{\text{0}}} = d_\text{h}([\boldsymbol{M}_{\psi},\boldsymbol{M}_{\zeta}],[\psi_{\text{0}},\zeta_{\text{0}}])$ \Comment{\eqref{eqn:dh}}
            \State $\text{Obtain (ascending) index for }\boldsymbol{d_{\text{0}}}, \boldsymbol{S}_{\text{I}}$
            \State $\boldsymbol{M}_{\text{Z}} = \boldsymbol{M}_{\text{Z}}(\boldsymbol{S}_{\text{I}})$
            \State $\boldsymbol{M}_{\psi} = \boldsymbol{M}_{\psi}(\boldsymbol{S}_{\text{I}}), \boldsymbol{M}_{\zeta} = \boldsymbol{M}_{\zeta}(\boldsymbol{S}_{\text{I}})$
            \State $z_{\text{k}}, v_{\text{k}} \gets\,\text{OrdinaryKriging}(\boldsymbol{M}_{\text{Z}}(\boldsymbol{S}_{m}), \boldsymbol{M}_{\psi}(\boldsymbol{S}_{m}), $
            $\hspace*{15.5em}\boldsymbol{M}_{\zeta}(\boldsymbol{S}_{m}), \psi_\text{0}, \zeta_\text{0})$%\Comment{Algorithm~\ref{alg:mc_kriging}}
            \If{$v_{\text{k}} < T_{v}$}
            \State $\boldsymbol{Z_{\text{}}}_{ij}, \boldsymbol{\Sigma_{\text{}}}_{ij} \gets z_{\text{k}}, v_{\text{k}}$
            \State $\textbf{I}_{\Omega ij} \gets 1$
            \EndIf
            \EndFor
            \EndFor
            \State \textbf{Global matrix completion:}
            %\State $\boldsymbol{\mathcal{I}} \gets ((-\alpha\Sigma_{\text{ij}},\alpha\Sigma_{\text{ij}}))$
            \State $\boldsymbol{Z_{\text{}}} \gets \text{NuclearNormMin}(\boldsymbol{Z_{\text{}}},\boldsymbol{\Sigma_{\text{}}},\textbf{I}_\Omega, \alpha)$\Comment{Algorithm~\ref{alg:mc_func}}
            %\State $\text{Complete matrix } \boldsymbol{Z_{\text{}}} \text{ as in \eqref{eqn:matrix_completion}}$
            \State \textbf{Prediction:}
            \State $\hat{Z}_{\text{mc}} \gets \text{interp2}(\boldsymbol{Z_{\text{}}}, \psi_\text{0}, \zeta_\text{0},``\text{nearest}")$
            \State $\textbf{Output:}\,\hat{Z}_{\text{mc}}$
            \EndProcedure
        \end{algorithmic}
\end{algorithm}

\subsection{Sparse Measurements Into Matrix}
\begin{figure*}
    \centering
    \begin{subfigure}{0.48\textwidth}
        \centering
        \includegraphics[width=\textwidth]{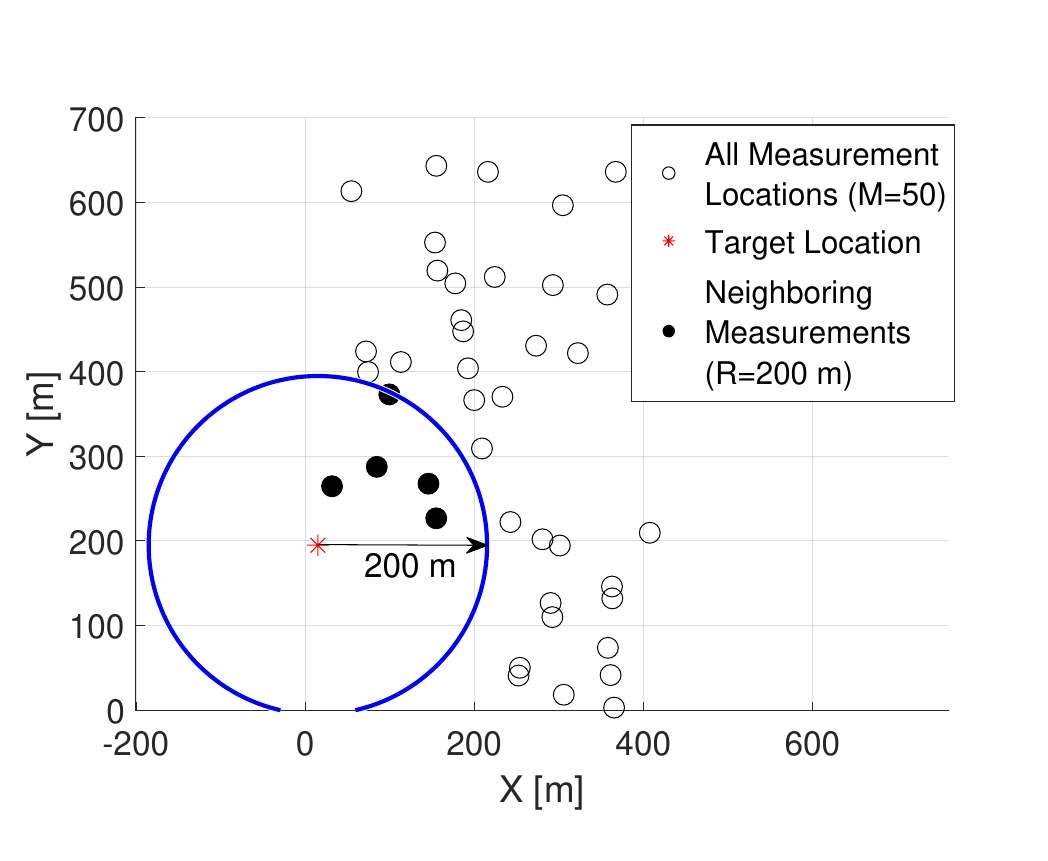}
        \caption{}
        \label{fig:sub1}
    \end{subfigure}
    \hfill
    \begin{subfigure}{0.48\textwidth}
        \centering
        \includegraphics[width=\textwidth]{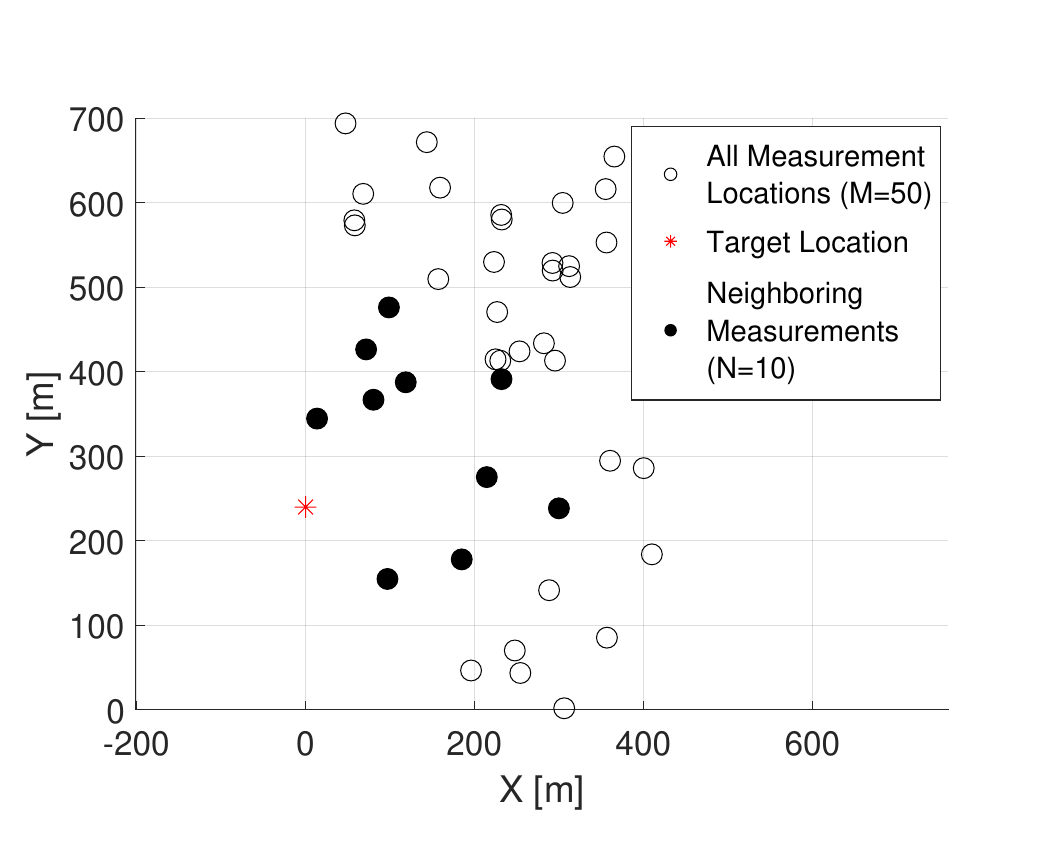}
        \caption{}
        \label{fig:sub2}
    \end{subfigure}
    \caption{Ways of choosing neighbors: (a) Fixed radius $R$, (b) Fixed number of training samples $N$ around an unknown point.}
    \label{fig:fixed_radius_and_points}
    \vspace{-1mm}
\end{figure*}

Matrix completion is an interpolation technique that takes a partially completed matrix as input and produces the full matrix as output. As this method is applicable only to grid data, the sparse RSRP measurements necessitate transformation into matrix form. To facilitate matrix completion for spectrum awareness, we obtained a bounded 2D region that encompasses the UAV trajectory in the horizontal plane. We then divide the region into small $5$~m by $5$~m square grids, where each grid corresponds to an entry of the matrix. Treating each grid's center coordinates as unknown points, RSRP values are estimated on those coordinates using neighboring measurements as shown in Fig.~\ref{fig:fixed_radius_and_points}(b). If the Kriging estimation on a grid center exhibits an estimated MSE lower than $T_v$, the estimation is considered reliable and matrix entry is considered known. The estimated MSE values associated with the known entries serve as the trust intervals for respective estimations. Let $\hat{Z}_{ij}$ denote the predicted RSRP of a grid center, and $\sigma^2_{ij}$ denote the corresponding estimation of MSE. For a Gaussian error distribution, $95\%$ confidence interval is approximately $(\hat{Z}_{ij}-2\sigma_{ij},\hat{Z}_{ij}+2\sigma_{ij})$.

\subsection{Constrained Nuclear Norm Minimization}
Constrained nuclear norm minimization~\cite{sun2022propagation} can be used for densifying a radio environment map. It requires a partially completed matrix that contains a set of known entries with expected MSEs for those entries. The unknown entries can be filled by applying the low-rank property of the propagation maps on the known entries. Let us denote the partially completed matrix $\textbf{H} \in \mathbb{R}^{N_\text{r} \times N_\text{c}}$, the matrix of estimated MSE ${\boldsymbol{\Sigma}}_\textbf{H} \equiv (\sigma_{ij}^2) \in \mathbb{R}^{N_\text{r} \times N_\text{c}}$ and the subset of indices with known measurement values $\Omega$. The completed matrix $\hat{\textbf{H}}$ can then be obtained through constrained nuclear norm minimization as follows~\cite{sun2022propagation}:
\begin{eqnarray}
\label{eqn:matrix_completion}
    \underset{\hat{\textbf{H}} \,\in\, \mathbb{R}^{N_\text{r} \times N_\text{c}}}{\operatorname{min}} & \|\hat{\textbf{H}}\|_* \nonumber \\
\text { subject to } & \hat{\text{H}}_{i j}-\text{H}_{i j} \in \mathcal{I}_{i j}, \quad \forall(i, j) \in {\Omega}
\end{eqnarray}
where $\mathcal{I}_{ij}$ is the trust interval for known $\hat{\text{H}}_{ij}$. We used an iterative approach to solve~\eqref{eqn:matrix_completion} as given in Algorithm~\ref{alg:mc_func}. 
It contains the $\text{FindNuclearNormApprox}(\cdot)$ function~\cite{Shabat2023} to approximately change the nuclear norm of a given matrix to a specified value. For the matrix completion process to produce a justifiable output, at least one entry of each column and each row must be initially known~\cite{sun2022propagation}.

In this work, we propose to obtain $\hat{\text{H}}_{ij}$ through Kriging interpolation and use corresponding $\sigma_{ij}$, obtained from ~\eqref{eqn:ordianry_Kriging_variance} or ~\eqref{eqn:simple_Kriging_variance} to construct a trust interval $\mathcal{I}_{ij}$ as follows:
\begin{equation}
\mathcal{I}_{ij} = \left(-\alpha\sigma_{ij}, \alpha\sigma_{ij}\right),
\end{equation}
where $\alpha$ is a tuning parameter that controls how much the algorithm is allowed to depart from the initially known values of $\textbf{H}$.
A very large value of $\alpha$ will ignore the initial values and so the norm of $\hat{\textbf{H}}$ will be nearly zero. On the other hand, if $\alpha$ is very small, the algorithm will approximately replicate the values from $\textbf{H}$ for the known entries. Thus, the rank of the $\hat{\textbf{H}}$ is implicitly characterized by $\alpha$.
It requires a thorough inspection of the grid distance, and correlation among neighboring cells to choose the value of~$\alpha$.
%balancing between the two extreme cases.
\begin{algorithm}[!t]
    \caption{Constrained nuclear norm minimization}
    \label{alg:mc_func}
        \begin{algorithmic}[1]
            \Procedure{$\textnormal{NuclearNormMin}$}{}%\Comment{The g.c.d. of a and b}
            \State $\textbf{Input:}\,\, \textbf{H} \!\in\! \mathbb{R}^{N_\text{r} \times N_\text{c}}, \boldsymbol{\Sigma} = (\sigma^2_{ij}) \in \mathbb{R}^{N_\text{r} \times N_\text{c}}, \textbf{I}_\Omega \in \{0,1\}^{N_\text{r} \times N_\text{c}}, \alpha$
            \State $\textbf{Initialize:}$
            \State $\hat{\textbf{H}} \gets \textbf{H}, \lambda \gets \|\textbf{H}\|_*$
            \State $\lambda_{\text{min}} \gets 0, \lambda_{\text{max}} \gets \|\textbf{H}\|_*, \delta_\lambda \gets \infty$
            %\State $ {\Omega} = \{(i, j) \in \mathbb{N} \times \mathbb{N} \mid i \leq N_\text{r}, j \leq N_\text{c}\}$
            \State $\textbf{Loop:}$
            \While{$\exists (i, j) \in {\Omega} \mid \abs{\hat{\text{H}}_{ij}-\text{H}_{i j}} \geq \alpha\sigma_{ij}\, \textbf{or}\, \delta_\lambda > T_{\lambda}$} 
            \State $\lambda_{\text{old}} \gets \lambda$
            \State $\lambda \gets (\lambda_{\text{min}}+\lambda_{\text{max}})/2$
            \For{$i\in\{1,\dots,n_{\text{iter}}\}$}
            \State $\hat{\textbf{H}} \gets \textbf{I}_\Omega\textbf{H} + (1-\textbf{I}_\Omega)\hat{\textbf{H}}$
            \State $\hat{\textbf{H}} \gets \text{FindNuclearNormApprox}(\hat{\textbf{H}}, \lambda)$\Comment{\cite{Shabat2023}}
            \EndFor
            \If {$\exists (i, j) \in {\Omega} \mid \abs{\hat{\text{H}}_{ij}-\text{H}_{i j}} \geq \alpha\sigma_{ij}$}
            $\lambda_{\text{min}} \gets \lambda$
            \Else
            $\,\,\lambda_{\text{max}} \gets \lambda$
            \EndIf
            $\delta_\lambda \gets \abs{\lambda-\lambda_{\text{old}}}$
            \EndWhile
            \State $\textbf{Output:}\,\hat{\textbf{H}}$
            \EndProcedure
        \end{algorithmic}
\end{algorithm}

\subsection{Prediction}
A 3D spatial location $\left(\psi_0, \zeta_0, h_0\right)$ consists of latitude, longitude, and height. Matrix completion uses the same height dataset for both training and testing. To predict the RSRP value of a new point, we apply 2D interpolation on the axes of latitude and longitude considering the grid centers of the matrix as the known points. The interpolation was conducted using MATLAB's $\text{interp2}(\cdot)$ function.

\section{Numerical Results}
\label{sec:num_results}
\subsection{Experimental Setup}
We varied the number of the known measurements $M$ at each UAV altitude between $50$ and $450$. For Kriging interpolation, we chose a neighborhood radius $R$ centered at the target location and used the known measurements inside the circle as the training samples for that location as depicted in Fig.~\ref{fig:fixed_radius_and_points}(a). In particular, $R$ was selected to be $70$~m, $100$~m, and $200$~m. For local interpolation in matrix completion, it is desirable to fill in as many entries as possible. However, the approach of Fig.~\ref{fig:fixed_radius_and_points}(a) does not guarantee to have neighbors around an arbitrary point. To obtain known measurements around a target point, we considered choosing the nearest $N$ known measurements around the target location as shown in Fig.~\ref{fig:fixed_radius_and_points}(b). This approach of interpolation
%with a fixed number of known samples 
is implemented within rows $16$-$20$ of Algorithm~\ref{alg:mc_algo}. In our experiment, $N$ was varied from $5$ to $20$. The hyperparameters of Algorithm~\ref{alg:mc_algo} were chosen as $T_v = 1000$, and $\alpha = 1$. The hyperparameters used in Algorithm~\ref{alg:mc_func} were $T_\lambda = 10$ and $n_{\text{iter}}=600$. Using $5$~m by $5$~m square grids we had a matrix of size $158$~by~$92$. Other than those for the 3D interpolation scenarios described in Section~\ref{sec:3d_interp}, we used the same height measurements for training and testing in all results.

\subsection{Matrix Completion}
\begin{figure}
\centerline{\includegraphics[width=\linewidth,trim={0.5cm 0.1cm 1cm 0.5cm},clip]{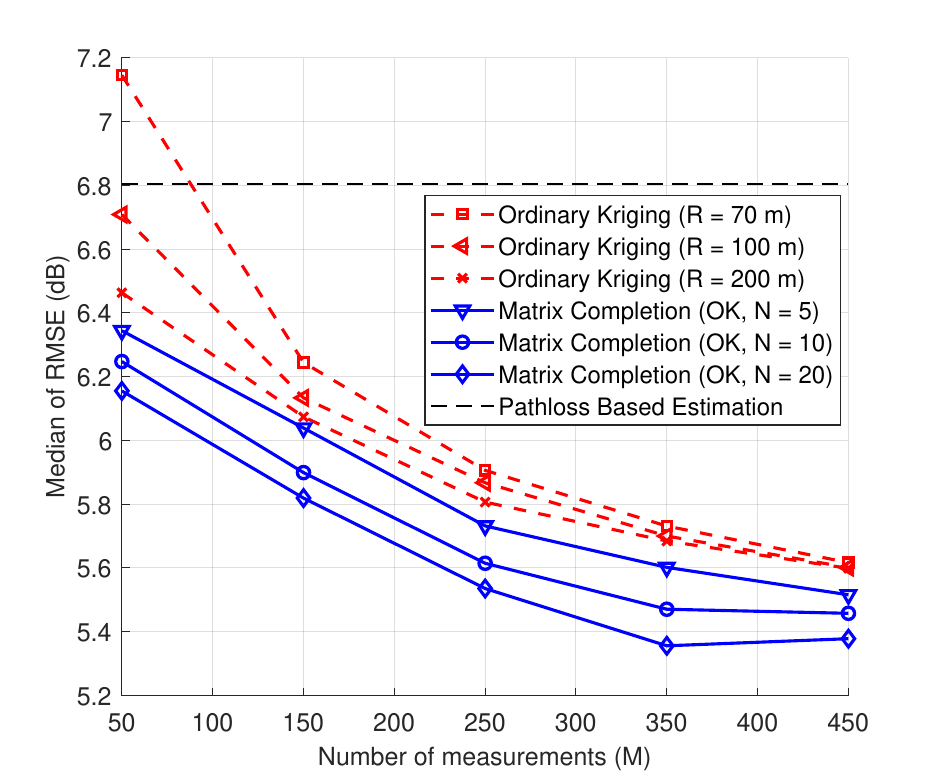}}
\caption{Performance comparison of ordinary Kriging and matrix completion for $110$~m height dataset. The best configuration of matrix completion ($N=20$) shows gain of $\sim\!0.2$~dB. over the best configuration of ordinary Kriging ($R=200$~m).}%compared to the best configuration of ordinary Kriging ($R=200$~m.)
\label{fig:matrix_completion_res}
\vspace{-4mm}
\end{figure}

Fig.~\ref{fig:matrix_completion_res} shows that the median of RMSE improves using matrix completion compared to ordinary Kriging. 
The local interpolation at each of the matrix entries has been computed using ordinary Kriging interpolation using the nearest $5$, $10$, or $20$ points. This result is for the height of $110$~m. Two key observations can be noted from this result. First, both ordinary Kriging and matrix completion boost performance with more data points around an unknown point through increased $R$ or increased $N$. Second, the best configuration of matrix completion has a gain of $\sim\!0.2$~dB compared to that of ordinary Kriging. Without global matrix completion, the results of ordinary Kriging and matrix completion are expected to be similar.
%as both follow Algorithm~\ref{alg:mc_kriging}.
However, when the matrix is adjusted in the process of matrix completion, the RSRP values that form a smoother pattern are prioritized, effectively suppressing the singular effects that might arise from measurement noise.
%\subsection{Combined Approach}

\subsection{Simple and Ordinary Kriging}
\begin{figure}
\centerline{\includegraphics[width=\linewidth,trim={0.5cm 0.1cm 1cm 0.5cm},clip]{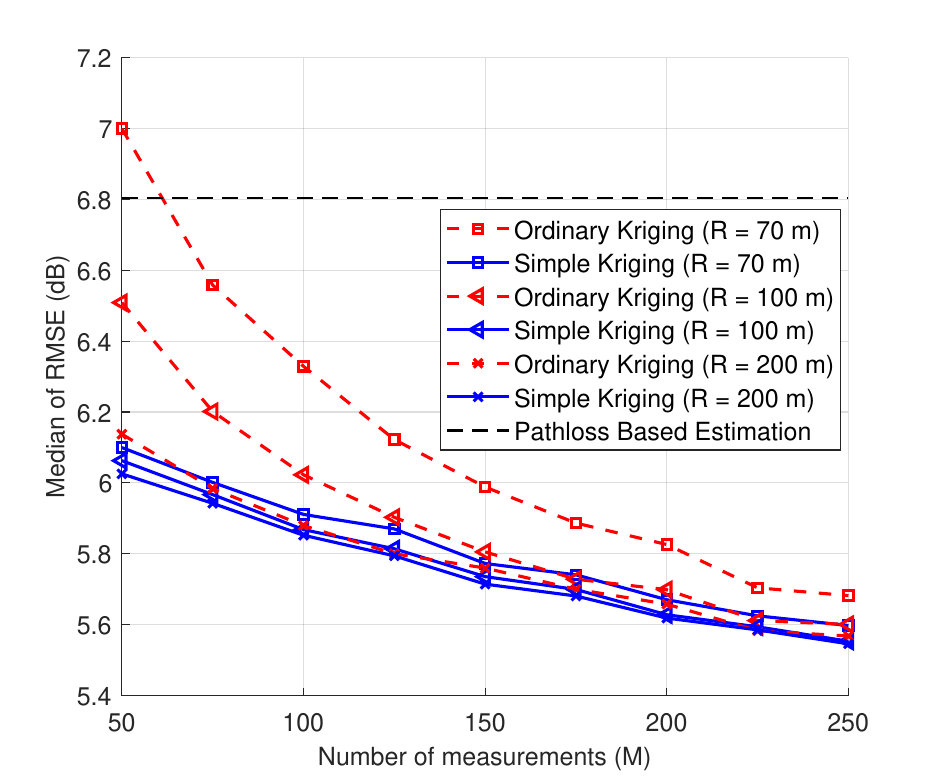}}
\caption{Performance of ordinary and simple Kriging for $70$~m height dataset. Simple Kriging outperforms ordinary Kriging more significantly for small $R$ and small $M$.} % for a small neighborhood radius $R$ and a small number of measurements $M$. outperforms ordinary Kriging for small $R$
\label{fig:simple_vs_ordinary}
\vspace{-4mm}
\end{figure}

Fig.~\ref{fig:simple_vs_ordinary} compares the performance of simple and ordinary Kriging as the number of measurement points increases from $50$ to $250$.
The largest performance gap occurs at the number of measurement points $M$ of 50 and the neighborhood radius $R$ of $70$~m. Simple Kriging outperforms ordinary Kriging by $0.8$~dB in this specific case. For other radii and the number of measurements, the performance gaps vary. The least difference is obtained for the $R$ of $200$~m. This result can be explained by comparing the number of unknown variables to solve between~\eqref{eq:ordinary_kirging_soln}~and~\eqref{eqn:simple_Kriging_soln}. Considering an equal number of known measurement points, ordinary Kriging has one extra unknown variable. When the number of known measurement points $N$ within a radius $R$ is small, the added model complexity for an extra variable is disproportionate to the number of known RSRPs. As a result, for smaller $R$ and smaller $M$, the negative impact of ordinary Kriging is more distinguishable.

\subsection{Trans-Gaussian Kriging}
\begin{figure}
\centerline{\includegraphics[width=\linewidth,trim={0.5cm 0.1cm 1cm 0.5cm},clip]{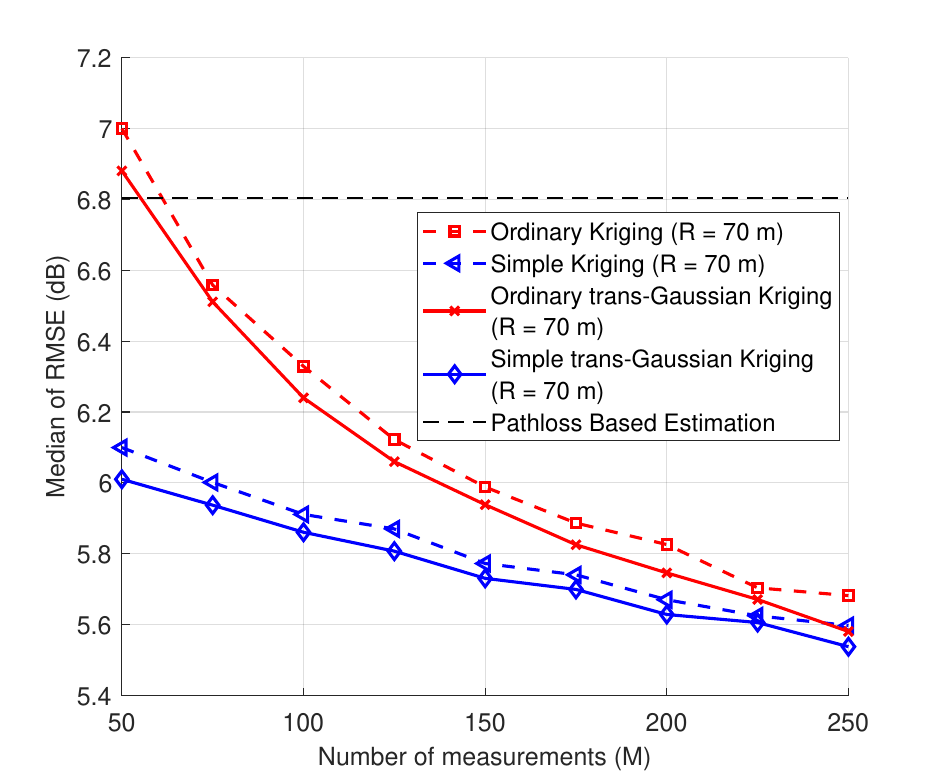}}
\caption{Comparison of ordinary Kriging, simple Kriging, and their trans-Gaussian variants for $70$~m height dataset. Trans-Gaussian is observed to yield performance improvement.} %by $\sim\!0.05$~dB
%The performance gap is marginal after the number of measurements reaches a certain number.}
\label{fig:simple_vs_ordinary_vs_gaussian}
\vspace{-3mm}
\end{figure}

Fig.~\ref{fig:simple_vs_ordinary_vs_gaussian} contains two plots of Fig.~\ref{fig:simple_vs_ordinary} that correspond to $70$~m neighborhood radius along with their trans-Gaussian Kriging result.
Trans-Gaussian Kriging improves the performance by $\sim\!0.05$~dB for both ordinary and simple Kriging. Similar to the case of Fig.~\ref{fig:simple_vs_ordinary}, we highlight the case of small $R$ ($70$~m) that produces a small number of measurements $N$. With fewer data points, the characteristics of the data points are more important, which benefits from the Gaussian distribution. Kriging formulates the prediction as a linear combination of the known measurements as in~\eqref{eqn:ordinary_kriging_inference}~and~\eqref{eqn:simple_kriging_inference}. When the number of measurements increases, a linear combination of many random variables is asymptotically nonlinear. On the other hand, with only a few variables, such a model is strictly linear, and hence it benefits from the  Gaussian distribution.
%When the number of measurement points crosses nearly 200, the curves seem to overlap each other and the performance gap becomes trivial.

\subsection{3D Interpolation}
\begin{figure*}
    \centering
    \begin{subfigure}{0.48\textwidth}
        \centering
        \includegraphics[width=\textwidth]{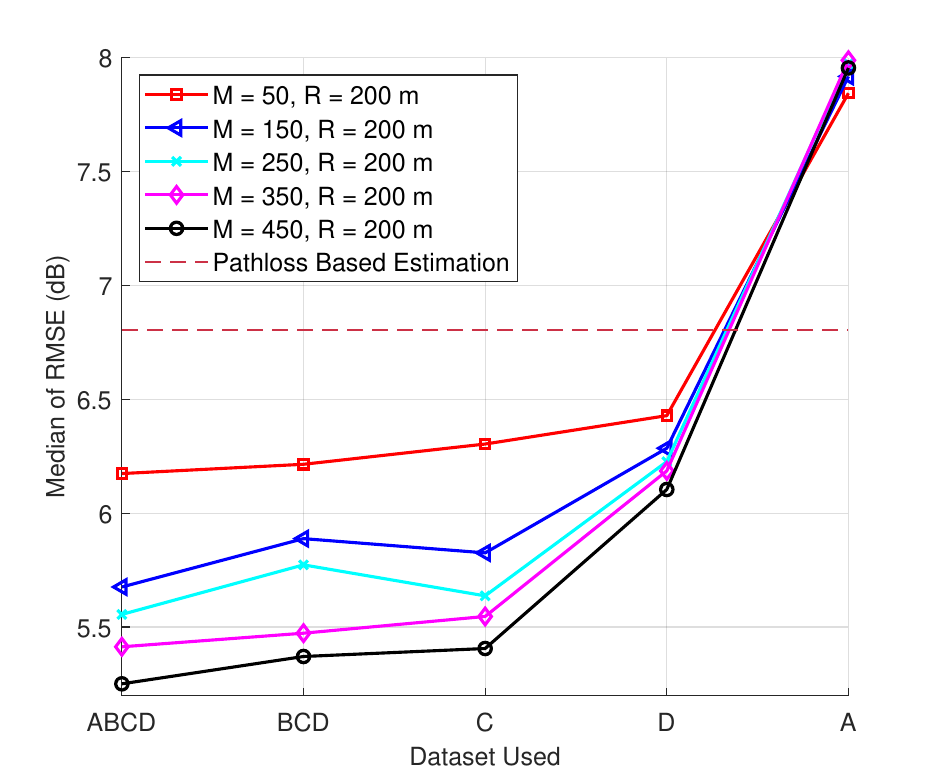}
        \caption{}
        \label{fig:sub111}
    \end{subfigure}
    \hfill
    \begin{subfigure}{0.48\textwidth}
        \centering
        \includegraphics[width=\textwidth]{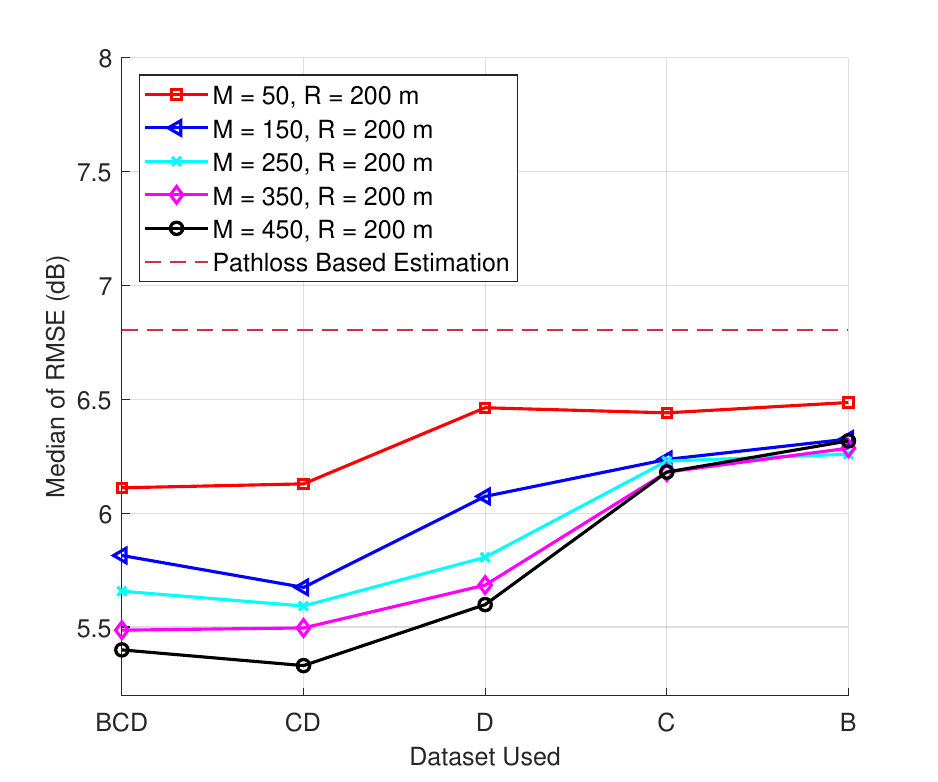}
        \caption{}
        \label{fig:sub222}
    \end{subfigure}
    \caption{Predicting RSRP values at (a) $90$~m height (Dataset C), (b) $110$~m height (Dataset D) from same height, another height, or multiple heights. Prediction using a dataset at $20$~m vertical distance shows comparable performance to the same height prediction, particularly for $M=50$. Combining the baseline dataset with that of other heights, such as $20$~m or $40$~m vertical distances can improve the performance.}
    \label{fig:3d_interp_combined}
    \vspace{-3mm}
\end{figure*}

In this subsection, we demonstrate how the RSRP estimation performance varies when we use the dataset from another height or multiple heights. 
Table~\ref{tab:dataset-names} contains a list of mixed datasets that contains the field measurement data either from a single height or multiple heights.
\begin{table}[]
\caption{Mixed dataset from field measurement data.}
\centering
\begin{tabular}{cc}
\hline
\textbf{Dataset Label} & \textbf{Field Measurement Heights} \\ \hline
A                & $50$~m                           \\ \hline
B                & $70$~m                           \\ \hline
C                & $90$~m                          \\ \hline
D                & $110$~m                          \\ \hline
CD               & $90$~m, $110$~m                    \\ \hline
BCD              & $70$~m, $90$~m, $110$~m              \\ \hline
ABCD             & $50$~m, $70$~m, $90$~m, $110$~m        \\ \hline
\end{tabular}
\label{tab:dataset-names}
\end{table}
Fig.~\ref{fig:3d_interp_combined} shows the interpolation performance at $90$~m (Dataset C) and $110$~m (Dataset D) heights using different training datasets listed in Table~\ref{tab:dataset-names}.
The plots show that prediction error using the data from locations that are vertically $20$~m or $40$~m away is larger than that of the same-height prediction, but still can perform better than pathloss-based estimation. However, predicting at $90$~m height from $50$~m data shows a poor performance. 
%The first two data columns in the left panels of Fig.~\ref{fig:3d_interp_combined} correspond to prediction using mixed data from heights. 
%For example, prediction using the dataset ABCD for the dataset C ($90$~m) utilizes data from the same height, two lower heights ($50$~m and $70$~m), and one upper height ($110$~m). 
The results show that utilizing the dataset at $20$~m vertical distance can provide comparable performance to the same-height result, especially when $M$ is low. When $M$ is increased, the performance at another height also increases, and simultaneously, the performance gap with the same height prediction also increases.
We have used an equal number of training samples $M$ at each UAV altitude, where $M$ varies from $50$ to $450$.

The first two data columns of Fig.~\ref{fig:3d_interp_combined} use a mixed dataset of multiple heights. Although this increases the number of training samples, we compare those in a row with single-height cases, as the dataset from another height is less likely to help.
The correlation among data points decays exponentially with the vertical distance as in~\eqref{eqn:corr_model}. The results show that using multiple heights data can improve performance.

%\subsection{Correlation model with both vertical and horizontal distance}

%\subsection{pathloss model combined with interpolation}

\section{conclusion}
\label{sec:conclusion}
In this study, using real-world 3D RSRP measurements collected at a UAV, we have conducted a comparative analysis of several variants of Kriging showing that simple Kriging and trans-Gaussian Kriging outperform ordinary Kriging, particularly for more challenging scenarios with limited measurements.
We have shown that the performance gap between simple and ordinary Kriging is large for small neighborhood radius.
We have also proposed a hybrid RSRP interpolation method using local Kriging with a fixed number of points and adjusting the matrix with nuclear norm minimization. The hybrid approach outperformed ordinary Kriging interpolation for both low and high numbers of measurements. Our results on 3D interpolation show that when the RSRP values are unavailable for a certain height, using the data from another height can provide improvements as long as the height difference is not large. We also demonstrated that the interpolation performance can be improved by combining data from multiple heights. These results provide important insights in facilitating spectrum awareness and sharing for real-world UAV scenarios.

\bibliographystyle{IEEEtran}
\bibliography{ref}

\end{document}